# Wafer-Scale Characterization of a Superconductor Integrated Circuit Fabrication Process, Using a Cryogenic Wafer Prober

Joshua T. West[1], Arthur Kurlej[2], Alex Wynn[2], Chad Rogers[1], Mark A. Gouker[2], and Sergey K. Tolpygo[2]

[1]High Precision Devices, Boulder, CO 80301 USA
[2]Lincoln Laboratory, Massachusetts Institute of Technology, Lexington, MA 02421 USA

*Abstract*— Using a fully automated cryogenic wafer prober, we measured superconductor fabrication process control monitors and simple integrated circuits on 200 mm wafers at 4.4 K, including SQIF-based magnetic field sensors, SQUID-based circuits for measuring inductors, Nb/Al-AlO$_x$/Nb Josephson junctions, test structures for measuring critical current of superconducting wires and vias, resistors, etc., to demonstrate the feasibility of using the system for characterizing niobium superconducting devices and integrated circuits on a wafer scale. Data on the wafer-scale distributions of the residual magnetic field, junction tunnel resistance, energy gap, inductance of multiple Nb layers, critical currents of interlayer vias are presented. Comparison with existing models is made. The wafers were fabricated in the SFQ5ee process, the fully planarized process with eight niobium layers and a layer of kinetic inductors, developed for superconductor electronics at MIT Lincoln Laboratory. The cryogenic wafer prober was developed at HPD/ FormFactor, Inc.

## I. Introduction

Fabrication process development of advanced nodes of superconductor electronics has reached stages when very large scale integration (VLSI) is possible and integrated digital circuits with millions of Josephson junctions and other components can be fabricated with high yield [1]-[6]. At the VLSI level, individual testing of digital circuits and circuit components at 4 K for determining fabrication yield, process parameters, parameter statistics and spreads across 200 mm, detecting rare outliers, studying repeatability from wafer-to wafer and from run-to run becomes extremely time consuming and expensive. Full-wafer probing for process control monitoring and circuit testing similar to the one used in semiconductor industry is required to further advance superconductor electronics for classical and neuromorphic computing, digital signal processing, quantum optimization, and quantum computing. This requires a fully-automated, cryogenic wafer prober operating at about 4 K and having very low levels of ambient magnetic fields and electromagnetic noise.

The interest in probing Josephson devices at cryogenic temperatures is long-standing, starting from the early development on 2-inch wafers [7] to a more recent work on probing resistance of a relatively large number of junctions on 150-mm wafers using a semi-automated probe station [8], and fully automated probing of junction resistance and critical current on a 1-cm chip scale [9].

The goals of this work were to demonstrate fully automatic testing of 200 mm wafers with superconductor integrated circuits and process control monitors (PCM) containing many tens of thousands of test structures, using a cryogenic wafer prober developed at Formfactor, Inc. [10]. The wafer fabrication was done at MIT Lincoln Laboratory (MIT LL). The SFQ 5ee fabrication process used has eight Nb superconducting layers, Nb/Al-AlO$_x$/Nb Josephson junctions with Josephson critical current density, $J_c$ of 100 µA/µm$^2$, a layer of Mo$_2$N kinetic inductors, and 2 Ω/sq thin film molybdenum resistors [11], [12].

Each PCM group comprised sixteen 5 mm × 5 mm chips printed 49 times on 200-mm wafers, using a 7 × 7 square grid with 22 mm step size. Each PCM group has about two thousand test structures which can be tested individually by contacting 100 µm × 100 µm contact pads with Nb/Pt/Au metallization, on 200 µm pitch. These test structures were designed for four-point measurements to allow for extraction of all basic process parameters and statistics, e.g., properties of Josephson junctions, inductors, resistors, critical currents of Nb wires and interlayer vias, etc.

## II. The Wafer Prober

### A. Design and Operation of the Cryogenic Wafer Prober

The cryogenic wafer prober is designed for testing 150 mm or 200 mm wafers at 4 K. The wafer is loaded using a standard Brooks wafer loading system [13] accepting standard 25-wafer

Manuscript receipt and acceptance dates will be inserted here. This material is based upon work supported by the Under Secretary of Defense for Research and Engineering under Air Force Contract No. FA8702-15-D-0001. *(Corresponding author: sergey.tolpygo@ll.mit.edu)*.

J. T. West (e-mail: Josh.West@formfactor.com) and C. Rogers are with High Precision Devices, Boulder, CO 80301 USA.

A. Kurlej (e-mail: Arthur.Kurlej@ll.mit.edu), A. Wynn (e-mail: alexander.wynn@ll.mit.edu), M.A. Gouker (gouker@ll.mit.edu), and S. K. Tolpygo (e-mail: sergey.tolpygo@ll.mit.edu) are with MIT Lincoln Laboratory, Lexington, MA 02421 USA

Color versions of one or more of the figures in this paper are available online at http://ieeexplore.ieee.org.

Digital Object Identifier will be inserted here upon acceptance.

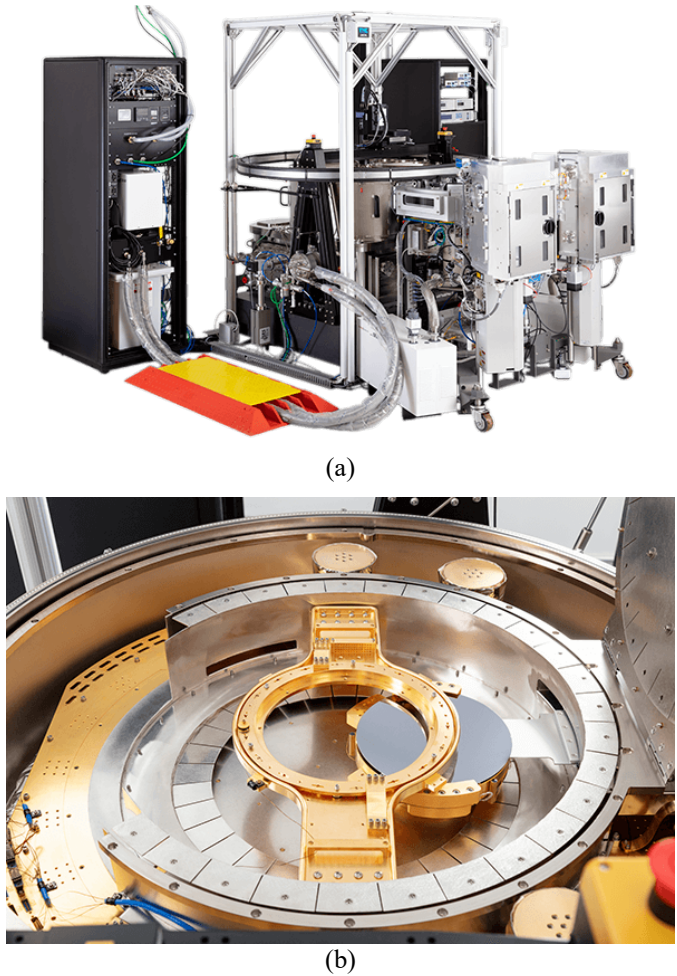

Fig. 1. (a) View of the full system, including cryogenics handling equipment rack, two load-lock chambers for loading cassettes with 200 mm wafer and 150 mm wafers, and coils for magnetic field canceling; (b) Close view inside the testing chamber during load sequence of a 200 mm wafer. Gray metal is mu-metal shielding.

cassettes. Wafer cassettes are loaded into the vacuum cassette elevator (VCE) that is evacuated prior to the wafer being picked up for loading into the cryogenic chamber. Wafers are picked from the VCE, pre-aligned, and then inserted directly onto the cryogenic chuck, while both the cryogenic chamber and the wafer handler at high vacuum, better than $10^{-6}$ Torr. The wafer is held to the chuck surface using a mechanical clamp mechanism. The chuck is actuated with $x$, $y$, $z$, and $\Theta$ (rotation about the $z$-axis) degrees of freedom using ultra-high-precision motion stages that operate within the vacuum space at 300 K.

Electrical contacts to the wafer are made using a replaceable probe card. For the series of tests described here, a DC probe card with 20 flexible tungsten probes (two rows of 10) was used. The pin-to-pin spacing was 200 µm and the two rows were separated by 200 µm (unstaggered) in order to match the design of the PCM chips. The probe card is mounted on a kinematic holder allowing for the adjustment of tip ($\theta_x$), tilt ($\theta_y$), and $z$-height such that the probe pins may be planarized to the surface of the wafer. Slotted mounting holes in the probe card allow for rotation ($\theta_z$) to align the probe pins to the axes of motion for the motion stages. The probe card holder is attached to the static cooling stage via flexible thermal links.

The chamber is wired with up to 500 twisted-pair DC wires and 56 RF coaxial lines with bandwidth up to 18 GHz. The cables are wired and heat sunk from 300 K to 4 K, allowing the operator to configure the probe card with the desired array of I/O signals. RF signals are terminated with SMA connectors at 300 K and SMP connectors at 4 K. DC wires are twisted pairs and terminated in D-Sub 25 connectors at 300 K and micro-D 25 connectors at 4 K.

Suppression of the Earth's magnetic field by more than two orders of magnitude is achieved with two layers of mu-metal shielding. A magnetic field is applied in all three axes using an external set of coils mounted on a cage structure as seen in Fig. 1(a). A passive mu-metal shield is mounted within the 4 K radiation shield for the second stage. The magnetic field within the prober is measured at three points around the perimeter of the wafer, using single axis fluxgate magnetometers [14].

A camera system with <3 µm resolution is used for imaging, navigation, automatic alignment routines, and precision die stepping. The camera is mounted on a three-axis positioning system (50 mm range) with a pneumatic lift stage for elevating the system when opening the probe chamber. The camera and positioning system are mounted on an ultra-rigid frame that allows the camera arm to be rotated out of the way for easy access to the probe chamber during probe card installation and maintenance.

Cooling of the wafer chuck and the static radiation shields is achieved with liquid cryogens. The system is pre-cooled with liquid nitrogen before it is purged and switched to liquid helium. Total cooldown time is less than 8 hours. Separate cooling loops are used for the static and motion stages. A binary gas analyzer monitors the helium purity of the exhaust gas and, if it meets a pre-set level, the exhaust may be diverted automatically to a helium reliquefication plant.

To accommodate testing of 150 mm and 200 mm wafers, the wafer chuck is mounted to the probe chamber with a single fastener locking mechanism that allows the entire wafer chuck to be replaced with one of a different size. The room temperature vacuum wafer handling system is equipped with two VCEs and two aligners (one for each wafer size), so no setup is required when changing wafer sizes.

A wafer $z$-height map is created after clamping to the chuck. Actual contact to the wafer is used to determine the contact height and a user-specified overdrive distance is input to ensure solid electrical contact with the device contact pads. A minimum separation distance is also set to ensure that no lateral motion is allowed unless the prober pins are at a safe distance from the wafer.

The wafer prober software handles all of the automation routines (pump down, cooldown, wafer loading/unloading, alignment, and die-to-die stepping). A wafer map is created showing the location of all testable dies. When an automated probing sequence is started, the prober moves to the first die and then sends a message to the user's test program that it is ready for measurement. The device is then tested, and a message is sent to the tester with a test/fail message and command to move to the next die or sub-die, if required. This sequence repeats until the entire wafer is tested. Die-to-die movement is typically less than 1 s. The pass/fail (or other binning criterion) is displayed in real time on the wafer map to give the user a visual representation of the status of the wafer testing.

## III. THE MEASUREMENTS

### A. Data Collection Methodology

Electrical measurements were specified using a user-built Matlab-based software "Measurement Manager" (MM). The software reads-in an xlsx-formatted file that specifies the location of each test structure on the wafer, the test to perform, and additional test-specific input fields. The MM software was run from a laptop computer, and interfaced with the test equipment through Ethernet and USB-GPIB controller.

All electrical connections were made by automated control of a Keithley 708B switch matrix with low-offset relays. Voltage measurements were made by a Keithley 2182a nanovoltmeter, and Yokogawa GS200 current sources were used to supply all input currents.

Automation of the full testing procedure included switching, motion control, and electrical measurements. Multiple types of tests, including $I$-$V$ measurements and SQUID- and SQIF modulation measurements were programmed in sequence via xlsx-formatted specification sheet, and full-wafer testing was performed with minimal need for human intervention.

In conventional chip-scale testing, a wafer is firstly diced into individual 5 mm × 5 mm chips, which are then wirebonded, loaded onto a test probe, and immersed in liquid helium for testing. Wafer dicing, cleaning, and chip picking alone accounts for several hours of overhead time caused by scheduling and execution, estimated at 4 hours per wafer. Delays due to wafer queuing for dicing availability can increase this time to several days in practice. Wire-bonding carries similar overheads, on the order of 1 hour per chip. Chip testing in a helium dunk probe is typically done in batches of 6 chips per load. Loading and cooldown takes about 15 minutes per 6 chips, although overheads due to testing in a dry cryostat are considerably higher. After cooldown, measurement time for a typical SQUID test structure, is about 1 minute.

Overall, conventional testing of 24 SQUID-based test circuits for extracting inductance of Nb layers at 22 different locations across a full wafer would require about 4 hours of dicing, plus 22 hours of wirebonding, plus 1 hour of loading, and about 5 hours of actual electrical testing, for a total of 32 hours. In practice, the calendar time may extend to a week or more due to queuing.

For a comparison, at wafer-scale testing, the loading is done once per wafer, with a cooldown time of about 10 minutes, measurement time is identical (performed by the same hardware), and chip-to-chip movement is the only additional contributor, and can be neglected in this example, at approximately 1 second per typical move. For the same 24 SQUID structures at 22 die locations, the total test time is approximately 5 hours, with the bulk of this limited by the actual electrical testing time. Hence, wafer-scale testing provides a dramatic reduction in the overhead time associated with chip-scale testing. Because of the overhead, the time savings with respect to the chip-level testing linearly increases with the number of chips requiring testing, while electric testing time in both methods is approximately the same.

### B. Magnetic Field Uniformity on Wafer Scale

Low residual magnetic field, below about 100 nT, is critical requirement to any system that is going to be used for testing superconductor digital electronics. For cryogenic test equipment intended to test diced chips, typically on the order of 1 cm × 1 cm, the requirement of low residual field is managed using a multilayer shield of high permeability metal such as cryoperm-10, annealed and assembled in low ambient field. For larger systems, achieving a sufficient aspect ratio of the shields would greatly increase the overall system size, weight and cost. Additionally, with locally varying magnetic elements such as the components within a motion stage, static cancellation of external fields is insufficient to manage local perturbations. With these considerations, the cryogenic wafer prober was designed to combine passive magnetic shields, as well as active cancellation from the set of external coils assembled on the cage structure visible in Fig. 1(a).

To evaluate the performance of the shielding system with respect to circuits under test, we measured SQIF magnetometers integrated into the test wafer as a component of the PCM chips. Each chip had 4 different nine-loop SQIFs in order to cover a wide range of possible magnetic fields. The SQIFs design was similar to the one described in [15]. The typical SQIF measurements (voltage-flux characteristics) are shown in Fig. 2(a), and the obtained wafermap of the $z$-component of magnetic field is shown in Fig. 2(b). There are easily detectable changes in the local magnetic field at the device under test location, which vary with position of the wafer chuck. We note that the field was always measured at the same global location with respect to the prober's testing chamber, right below the tips of the probe card, but using SQIFs on different chips brought under the tips by the chuck movements. Hence, the wafermap in Fig. 2(b) does not represent distribution of the field within the testing chamber (which was not measured) but reflects changes in the field caused by a relative position of the chuck with respect to the chamber walls.

With active compensation on, all test sites showed a consistent offset field (between measured by the fluxgate magnetometers and the SQIFs) of about 80 nT. Higher fields were observed at the edge of the wafer when the center of the chuck is closer to walls of the test chamber and mu-metal shields in Fig. 1(b), particularly in the lower-left corner; see Fig. 2(b). The reasons for this need further investigation. Overall the measured on-wafer fields correlate with the system-level measurements using fluxgate magnetometers used by the prober's magnetic field cancellation system.

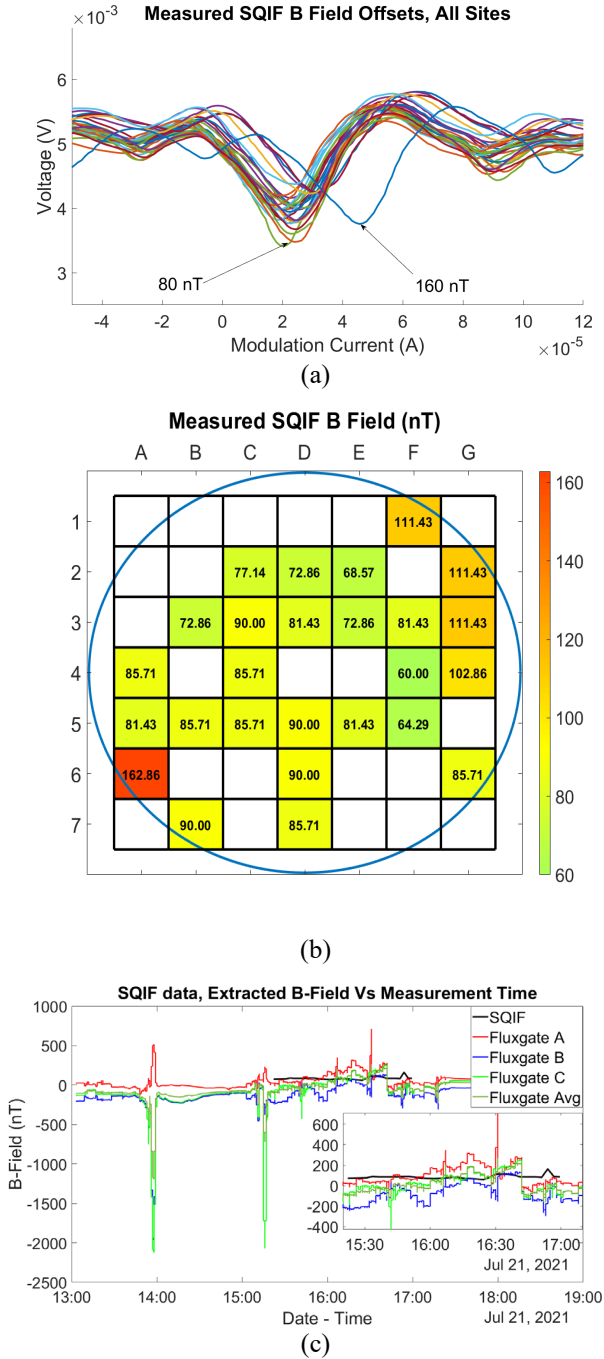

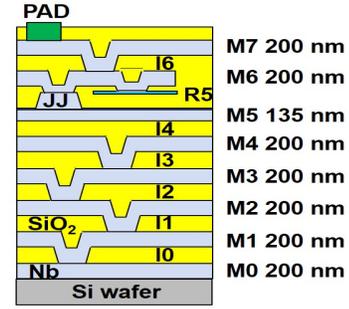

Fig. 3. Simplified cross section of the SFQ5ee process used for wafer-scale characterization. The nominal thickness of all Nb layers is 200 nm, except the junction base electrode layer M5 which is 135 nm after a light anodization of its surface used to encapsulate etched JJs into anodic oxide. The nominal thickness of interlayer dielectrics I0 through I5 is 200 nm, I6 is 280 nm. Josephson junctions are formed between Nb layer M5 and the counter electrode labeled JJ, using an in-situ deposition and oxidation of a thin Al layer. After etching the JJs, this Al layer is converted into anodic oxide during the M5 anodization (shown as a black line on the M5 surface). Resistor layer R5 (40 nm thick) is used to form shunt and bias resistors. The cross section shows a composite via from layer M0 to layer M4, which is referred to as via I0I1I2I3 and presents four staggered individual vias I0, I1, I2, I3.

Stability of the field cancelation system during a 6-hour period was tested and demonstrated in Fig. 3(c). After the cooldown, a one-time measurement of magnetic field in the system was made by the fluxgate magnetometers to setup currents in the prober's compensation coils and cancel the measured field. These currents then were held for an extended period of time. Fig. 3(c) shows the field components measured by the three-axes fluxgates sensor (blue, red and green) and the average field (olive). SQIF measurements are shown by the black curve. Spikes in the magnetic field at 14:00 and 15:15 correspond to loading wafers on the cold chuck, with the SQIF wafer (wafer # SFQ511) loaded at 15:15. These spikes in the fluxgate sensor output during the wafer load are likely caused by temperature gradients the sensor is experiencing during this time (going from 4 K to ~40 K, and back in minutes). The system magnetic environment became especially quiet after approximately 17:30, roughly corresponding to the end of the working hours, when all heavy machinery and equipment at the system location were switched off. No attempt was made to use the SQIF data to tune the magnetic cancellation system for optimal performance. Also, no active feedback was provided during the measurements. However, this all could be done in the future.

Fig. 2. (a) Flux-voltage characteristics of SQIF magnetometers measured cross-wafer. The majority of sites showed a consistent residual field of about 80 nT. On-chip SQIFs were modulated using integrated with each SQIF modulation coils. Off-set of the main voltage minimum from the zero modulation current is the measure of the residual magnetic field at the SQIF location on the wafer. (b) Wafer map of the magnetic field measured by the SQIF magnetometers located on different chips. The field was measured at the same global location (under the probe card tips). Changes in the observed field are caused by changes in the relative position of the wafer chuck and the testing chamber walls. (c) Long-term stability of the magnetic field cancelation system using three-axes fluxgate sensors (red, green, blue, and olive showing the average), and by z-axis SQIF magnetometers (black). Spikes at 14:00 and 15:15 correspond to changing wafers on the cold chuck of the prober. The SFQ511 wafer with the SQIF sensors was loaded at 15:15. Inset shows a zoom into the time period of the SQIF wafer testing from 15:20 to 17:00.

### C. Wafer Fabrication and the SFQ5ee Process Details

We used wafers with superconductor circuits and process control monitors (PCMs) fabricated in the SFQ5ee process developed for superconductor electronics at MIT Lincoln Laboratory. It is a planarized process with eight Nb layers and Nb/Al-AlOx/Nb Josephson junctions [12]. A schematic cross section of the process is shown in Fig. 3. Niobium layers are labeled M0, M1, etc. The Josephson junctions are formed between the layer M5 and the counter electrode labeled JJ in Fig. 3. Resistor layer R5 is used to shunt Josephson junctions and form circuit resistors. A thin layer of high kinetic inductance material Mo$_2$N placed below Nb layer M0 is not shown and was not used in the measurements. Vias between Nb layers are formed by etching contact holes in the planarized $SiO_2$ interlayer dielectric and filling them with Nb of the next layer. Planarization of all dielectric layers I0, I1, etc. is done by chemical mechanical polishing. Vias in dielectric layers are also labeled as I0, I1, etc. Thicknesses of all layers are given in Fig. 3. More process details can be found in [6], [12]; see also Table 1 in [16].

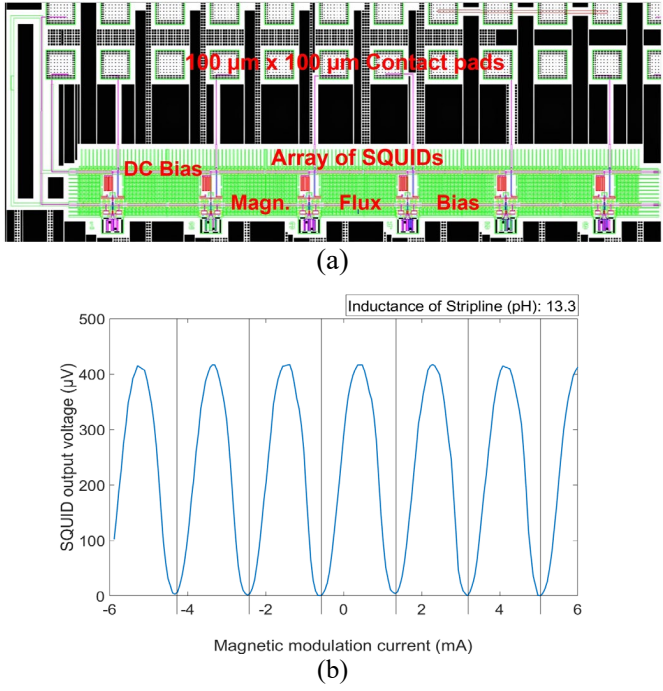

Fig. 4. (a) Layout of a 6-SQUID array, a fragment of the integrated circuit used for inductance measurements, requiring $N + 2$ contact pads to measure inductance of $N$ inductors on PCM chips. Using the wafer prober, these circuits were measured across the wafer. (b) The typical voltage vs. magnetic bias current characteristics of one of the SQUIDs (#12) in the integrated circuit, corresponding to a stripline inductor with signal wire formed on Nb layer M6 between two ground plane niobium layers M4 and M7, stripline M6aM4bM.

### D. Testing Microstrip and Stripline Inductors

With the ambient field in the system well characterized, the system was used for rapid evaluation of several important process parameters. Inductors are the second most important component, after Josephson junctions, for all superconductor electronic circuits. However, they are one of the very few circuit elements for which room-temperature measurements cannot provide input regarding parameter targeting or variation. There are nine superconducting layers in the SFQ5ee process, allowing circuit designers to form a few dozen of different types of inductors comprising two superconducting layers (microstrip types) or three layers (striplines). Knowledge and control of linear inductance of multiple combinations of niobium layers and their dependence on the linewidth are essential for the stable fabrication process and circuit design. Therefore, ability to measure large numbers of inductor structures at 4 K is a critical enabling capability.

To measure inductors we used integrated circuits which require $N + 2$ contact pads to extract inductance of $N$ inductors from the periods of voltage-flux characteristics of differential SQUIDs fed by a common modulation current; the SQUID volt-age is measured with respect to the common ground [17]. Results of the extensive chip-level measurements using this approach were given in [18], [19]. A fragment of the typical layout of the inductance test circuit is presented in Fig. 4a, showing an array of six SQUIDs, the DC bias and magnetic flux bias rails.

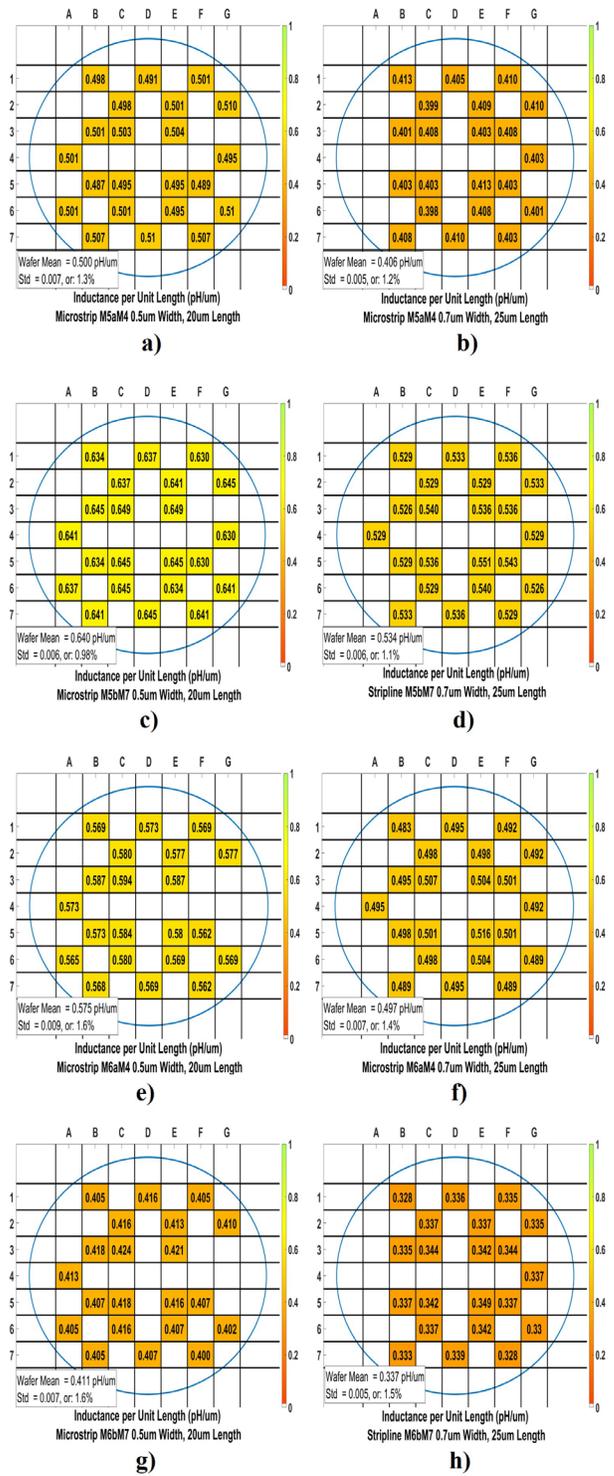

Fig. 5. Wafermaps of inductance per unit length of microstrips M5aM4 (a), (b); microstrips M5bM7 (c), (d); microstrips M6aM4 (e), (f); and microstrips MbM7 (g), (h) with linewidths of 0.5 µm and 0.7 µm, fabricated on 200-mm wafers in the SFQ5ee fabrication process. Niobium layers M4 and M7 are the ground planes. The wafer mean and "standard deviation" (second moment of the distribution) are given. Collecting this amount of information using the chip-level testing would take at least several days.

TABLE I
EXPECTED INDUCTANCE AND DEVIATION OF THE MEASURED WAFER MEAN VALUES

| Inductor | $d_1$ (nm) | $L_l$ (pH/µm) $w=0.5$ µm | $\frac{\Delta L_l}{L_l^{theor}}$ (%) [a] | $L_l$ (pH/µm) $w=0.7$ µm | $\frac{\Delta L_l}{L_l^{theor}}$ (%) |
|---|---|---|---|---|---|
| M5aM4 | 200 | 0.478 | 4.6 | 0.383 | 6.0 |
| M5bM7 | 680 | 0.645 | −0.8 | 0.549 | −2.6 |
| M6aM4 | 615 | 0.569 | 1.0 | 0.490 | 1.4 |
| M6bM7 | 200 | 0.427 | −3.7 | 0.350 | −3.7 |
| M5aM4bM7 | 200 | 0.4472 | 1.5 | 0.3556 | 2.1 |
| M6aM4bM7 | 615 | 0.3856 | n/a | 0.3118 | n/a |

[a] $\Delta L_l = \langle L_l^{meas} \rangle - L_l^{theor}$, where $\langle L_l^{meas} \rangle$ is the measured wafer mean value of inductance per unit length, and $L_l^{theor}$ is the expected value from (1) with nominal process parameters. Penetration depth $\lambda = \lambda_1 = 90$ nm was used in (1) and (2).

The typical voltage-flux (magnetic bias current) characteristic of one of the SQUIDs (#12) in the integrated circuit is shown in Fig. 4b. The $V$ vs $\Phi$ characteristics measured on the prober allowed for an accurate extraction of the modulation period and were very similar to those measured in the chip-level test set-up. We were mostly interested in inductors formed on the layers closest to the layer of Josephson junctions: microstrips with signal traces on layers M5 or M6 above M4 ground plane, labeled respectively, M5aM4 and M6aM4; striplines with signal traces on layers M5 or M6 above M4 and below M7 ground planes, respectively, M5aM4bM7 and M6aM4bM7 in the terminology of [18], [19].

The wafermaps of the extracted inductance of different microstrip inductors with signal traces on the layers M5 and M6 are shown in Fig. 5 for two linewidths $w = 0.50$ µm and 0.70 µm. Because these linewidth are relatively large, significantly larger than the practical resolution limit of process, the variation of inductance on the wafer is not related to the variation of the linewidth across the wafer but rather represents a cumulative result of the variation in the dielectric thickness, $d_1$ (between the signal trace and the ground plane) and in the magnetic field penetration depth, $\lambda$. This follows from the expression for microstrip inductance derived in [19]

$$L_l = \frac{\mu\mu_0}{4\pi} ln\left[1 + \frac{4(d_1+t/2+\lambda)^2}{0.2235^2(w+t)^2}\right] + \frac{\mu_0\lambda_1^2}{wt}, \quad (1)$$

where the first term is magnetic inductance and the second term is kinetic inductance, $t$ is the signal trace thickness that is 135 nm for the M5 layer and 200 nm for all other layers; $\mu$ and $\mu_0$ are the relative and vacuum permeability. We use the same penetration depth in Nb of the signal traces, $\lambda_1$ as in the ground planes, $\lambda = \lambda_1 = 90$ nm to be consistent with the data in [18], [19].

The expected inductances from (1) are shown in Table I for the nominal thickness of layers in the SFQ5ee process and $\lambda = 90$ nm. The relative difference of the measured wafer-mean values $\langle L_l^{meas} \rangle$ in Fig. 5 and the theoretical values is also given in Table I for the two linewidths. The percent difference is a factor two smaller for inductors M6aM4 and M5bM7 with the target $d_1 > 600$ nm than for inductors M5aM4 and M6bM7 with $d_1 = 200$ nm. This is easy to explain because the thick dielectric between M6 and M4 layers and between M5 and M7 layers is produced as a result of depositing and chemical-mechanical polishing of multiple layers of dielectric, whereas the thin, 200-nm dielectric is deposited and polished as a single layer. As a results of multiple independent depositions and polishing steps, the relative thickness deviation of the composite layers of dielectric from the target thickness is less than of the individual layers comprising them.

To access the wafer-scale uniformity, we compared the average inductance of the four central dies (C3, C5, E3, E5) and of the wafer periphery dies (B1, B7, D1, D7, F1, F7). More signal traces on the M5 layers there is no statistical difference between the center of the wafer and the periphery. The inductance difference is less than 1%, which is lower than the error of the measurements, estimated as 1.5% in [18], and consistent with the very low standard deviations given in Fig. 5a-d.

On the other hand, for signal traces on the M6 layer, there is a statistically significant difference of about 3.3% between the central dies and the dies at the periphery of the wafer. Regardless of the linewidth, inductance in the central part of the wafer is higher. This wafer-scale variation of circuit inductors is similar to the one reported in [18], [19] based on the chip-level measurements. Because of this systematic center-to-edge variation, the distribution of inductance values is not Gaussian and the given "standard deviation" simply represents the second moment of this distribution.

Wafermaps of the inductance per unit length of stripline inductors M5aM4bM7 are shown in Fig. 6. Inductance of the striplines is given in [19] by

$$L_l = \frac{\mu\mu_0}{4\pi} ln\left(1 + \frac{sin^2\frac{\pi(d_1+0.5t+\lambda)}{(H+2\lambda)}}{sinh^2\frac{\pi r_{eq}}{2(H+2\lambda)}}\right) + \frac{\mu_0}{8\pi} + \frac{\mu_0\lambda_1^2}{wt}, \quad (2)$$

where $r_{eq}$ is the equivalent radius of a cylindrical wire producing the same far field as the rectangular strip; see the expression

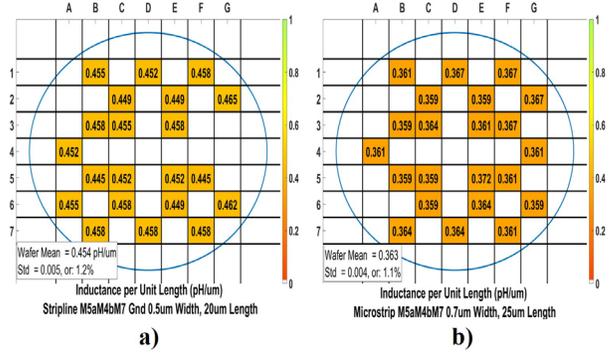

Fig. 6. Wafermap of inductance per unit length of striplines M5 above M4 and below M7 ground planes (M5aM4bM7). The nominal distance between the ground planes, $H$ is 1015 nm, the thickness of the M5 layer is 135 nm.

(23) in [19]. H is the distance between the two ground planes. In the SFQ5ee process the nominal distance between M4 and M7 ground planes is 1015 nm; see [19]. For the M5 strips with $w = 0.5$ μm and 0.7 μm, the $r_{eq}$ is, respectively, 0.1807 μm and 0.2337 μm. The first two terms in (2) is magnetic inductance associated with magnetic field in and around the signal wire and the third term is the kinetic inductance of the supercurrent. We see again a very good agreement between the measured and the expected inductance values and a very uniform distribution of the stripline inductance across the wafer.

*E. Superconducting Vias Between Nb layers*

Vias between superconducting layers are the second component for which no information about their critical current in the superconducting state can be obtained based on room-temperature characterization. Critical current of vias between all metal layers was measured on the cryogenic prober. The measurements were done using chains of 1848 vias connected in series. Since voltage is measured across the entire chain, this method measures the smallest critical current of vias in the chain, $I_{c,via}$ but does not characterize the distribution of the critical currents within the chain. This distribution is actually not important because this is a pass/fail test for the critical current which must exceed a certain level, typically 15 mA per square via with size of 700 nm.

Six 1848-via chains per site were measured for each via type, and the mean critical current was calculated. The typical wafermaps for all types of vias show radially symmetric distribution of the mean critical current: the critical currents are higher near the center of the wafer and lower near the edges; see Fig. 7. This distribution agrees and supplements the limited datasets collected previously by dicing and wire-bonding individual chips.

*F. Josephson Junctions*

Tunnel barrier resistance of Nb/Al-AlO$_x$/Nb junctions at room temperatures, $R_{300}$ is directly related to the tunnel barrier resistance at cryogenic temperatures and to the critical current of Josephson junctions via an $I_c R_N$ product, a parameter specific for the fabrication process, where $I_c$ is the critical current and $R_N$ is the normal-state resistance (above the critical temperature) of the Josephson junctions. This allows extracting the basic characteristics of JJs, e.g., resistance wafermaps and statistics on a very large number of JJs from room-temperature measurements using a conventional wafer prober; see [6], [10].

Measurements of JJs of different sizes on the cryogenic wafer prober allow us to obtain current-voltage (*I-V*) characteristics of the JJs to

a) determine the critical current $I_c$, normal state resistance, $R_N$, $I_c R_N$ product, the Josephson critical current density $J_c$, and also the gap voltage $V_g = 2\Delta/e$, where $\Delta$ is the energy gap in the spectrum of single-particle excitations in superconductors [21] and $e$ electron charge;

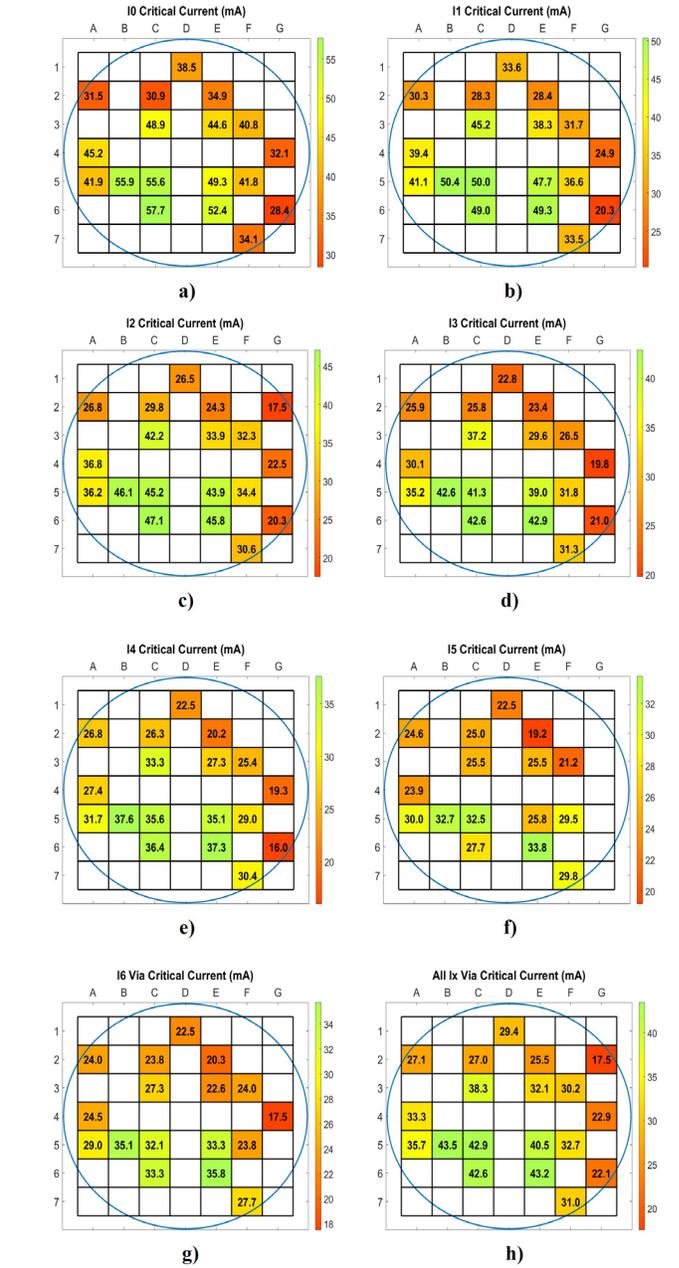

Fig. 7. (a) Wafermaps of the mean critical current of strings of 1848 vias, 0.7 μm square between different niobium layers in the SFQ5ee process: a) between layers M0 and M1, called I0 vias; b) between Nb layers M1 and M2, called I1; c) between Nb layers M2 and M3, called I2; d) between Nb layers M3 and M4, called I3; e) between Nb layers M4 and M5, called I4; f) between Nb layers M5 and M6, called I5; g) between Nb layers M6 and M7, called I6; h) composite via from the first Nb layer M0 to the last niobium layer M7, called All_Ix Via. Six strings of 1848 vias of each type were measured at each location to calculate the mean value.

b) check the consistency of the $R_{300}/R_N$ ratio across the wafer;

c) measure the subgap resistance in superconducting state, $R_{sg}$, an important characteristic of the tunnel barrier quality, which cannot be measured at 300 K;

In this work, to obtain the wafermaps, we measured six Josephson junctions of each of the four different sizes: 500 nm, 700 nm, 1000 nm, and 1400 nm in diameter. Fig. 8 shows *I-V* characteristics of JJs with diameters 1000 nm and 1400 nm, measured at 21 locations across the wafer. One can see that the Josephson critical current of the individual junctions was significantly suppressed, apparently because the test set-up did not have sufficient electromagnetic filtering right at the entrance of wiring into the prober. This was not important for testing circuits for inductance extraction (Fig. 4), which showed full critical currents, because those chips include integrated on-chip filters. Hence adding either cold filters to the wiring of the wafer prober cards or external filters would be advisable.

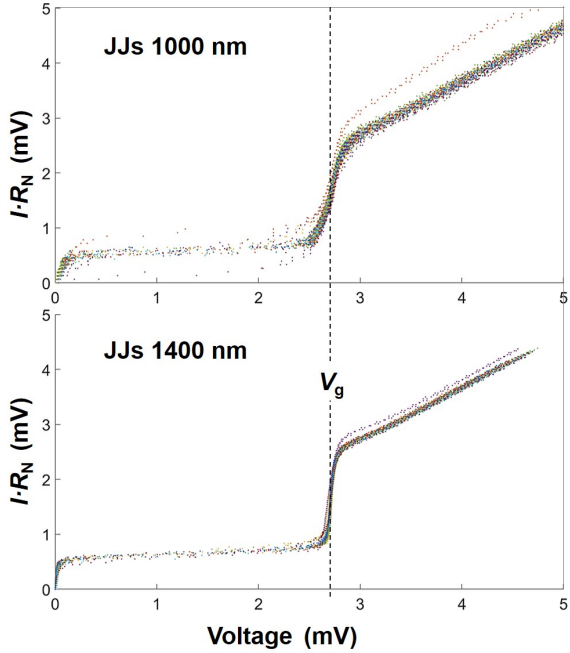

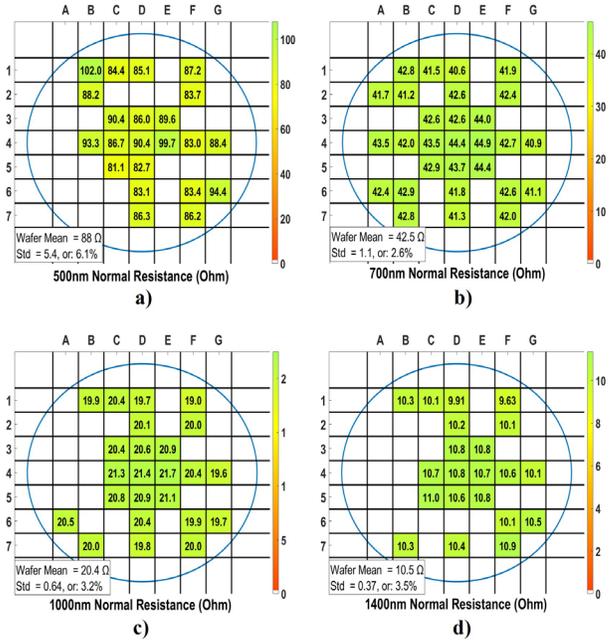

Fig. 8. Superimposed *I-V* characteristics of Josephson junctions measured at 21 locations across the 200 mm wafer: circular junctions with diameter 1000 nm (top panel) and 1400 nm (bottom panel). Six junctions of each size were measured at each location. Josephson critical current of the individual JJs was suppressed because of insufficient electromagnetic filtering in the test set-up.

Fig. 9. Wafermaps of the normal-state resistance, $R_N$ of circular Josephson junctions at 4.2 K measured on the 200-mm wafer (the same as in Fig. 8) from the slope of *I-V* characteristics above the gap voltage for JJs with diameters: a) 500 nm; b) 700 nm, c) 1000 nm; d) 1400 nm. The data are the mean resistance of six JJs measured at each location. The shown standard deviations should be viewed simply as a second moment (normalized to the mean) of the wafer-scale (non-Gaussian) distribution of properties of Nb/Al-AlO$_x$/Nb tri-layers on 200-mm wafers in the SFQ5ee fabrication process.

Wafermaps of the average normal state resistance determined as an average slope of *I-V* characteristics above 3.5 mV are shown in Fig. 9 for the four sizes of JJs. We can see that, independently of the JJ size, all wafermaps show the same pattern reflecting the wafer-scale properties of Nb/Al-AlO$_x$/Nb trilayers in the SFQ5ee fabrication process at MIT Lincoln Lab. The pattern is somewhat similar to the one observed for inductors and vias; see Fig. 5 and Fig. 7. The $R_N$ of JJs is higher (the Josephson critical current density $J_c$ is lower) in the central part of the wafer and reduces ($J_c$ increases) going towards the edges of the wafer, demonstrating nearly cylindrical symmetry. This trend is especially clearly seen for 1000-nm and 1400-nm JJs due their small on-chip variation (standard deviation of about 1%) which does not mask the global wafer-scale variation of the trilayer properties. The typical center to edge change in the $R_N$ and $J_c$ is about 9%. This is fully consistent with the wafermaps obtained in the room-temperature measurements of $R_{300}$ [6], [11], [20].

Similar data for another wafer fabricated in a completely different fabrication process run are presented in Fig. 10 to characterize the run-to-run repeatability of resistance (and critical current density targeting) and reproducibility of the wafer-scale distribution of JJ properties.

Another important characteristic of the process uniformity is the gap voltage $V_g$ of the Josephson tunnel junctions, which characterizes local properties of niobium base and counter electrodes of the junctions. Wafersmaps of the gap voltage for circular Nb/Al-AlO$_x$/Nb junctions with diameters 1000 nm and 1400 nm are shown in Fig. 8 for the junctions on the same wafer as in Fig. 6. The gap voltage was taken as the voltage corresponding to the maximum of the derivative $dI/dV$ of the *I-V* characteristics. We see a very good uniformity of $V_g$ indicating uniformity of the superconducting gap in Nb electrodes of the junctions.

## IV. DISCUSSION

### A. Josephson junctions

All obtained distributions of the fabrication process parameters have nearly the same wafermaps showing elevated values in the central part of the wafer, decreasing values towards the wafer edge, and nearly cylindrical symmetry. This is certainly caused by the processing equipment used which all has similar distribution of plasma properties, mainly, a dc magnetron sputtering of niobium and high density plasma etching of Nb and SiO$_2$ interlayer dielectric. For instance, distribution of $R_N$ in Fig. 9 and Fig. 10 and inverse to it distribution of $J_c$, since $J_c =$ $\alpha V_g/(R_N A)$, was attributed in [6] to changes in the work function of niobium caused by the intrinsic stress distribution in Nb films; here $A$ is the junction area, and $\alpha$ is a dimensionless parameter specific for the fabrication process. The film stress is slightly compressive in the central area of the wafers and increases to strongly compressive towards wafer edges due to a specific pattern of magnetron sputtering gun with rotating magnets in Endura PVD system used for depositing Nb films.

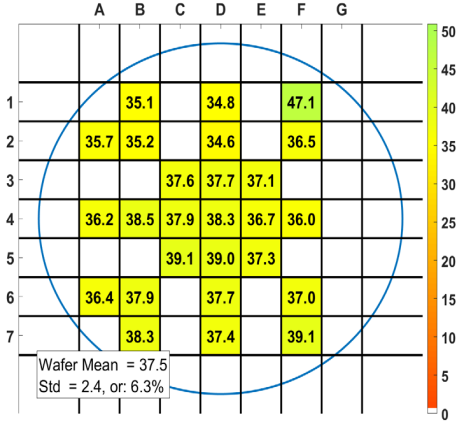

Fig. 10. Wafermap of the normal-state resistance, $R_N$ of 700-nm circular Josephson junctions at 4.2 K measured on a different wafer, SFQ511, fabricated in a different fabrication run than the wafer in Fig. 9. The wafermap is quite similar to the ones in Fig. 9, indicating good repeatability of Nb/Al-AlO$_x$/Nb trilayer properties from run to run.

Current-voltage ($I$-$V$) characteristics of the junctions fabricated in the SFQ5ee process were studied in [22]. On chips submerged in liquid helium ($T = 4.24$ K), the junctions show the gap voltage of 2.78 mV; see Fig. 3 in [22]. The smaller values of the $V_g$ in Fig. 11 and the apparent reduction of the $V_g$ in larger JJs in Fig. 11(b) with respect to the smaller JJs in Fig. 11(a) is an artifact related to internal Joule heating inside the junctions. When a JJ switches from the superconducting state to the gap voltage, an influx of nonequilibrium quasiparticles with energy $2\Delta$ starts. Their energy is eventually converted into Joule heating. Power dissipation $P = IV_g = \pi d^2 J_c V_g/4$ increases with the junction diameter as $d^2$. At the same time, the heat removal from the junction is proportional mainly to the effective area of Nb wires connecting the junction, $(d + 2s)l_{spr}$, because most of the heat conduction occurs along niobium wires due much higher thermal conductivity of Nb than of the surrounding amorphous SiO$_2$, where $s = 500$ nm is the surround of the JJs by Nb wiring used on this test chip and $l_{spr}$ is the heat spreading length along Nb wires. As a results, the increase in internal temperature of the junctions is proportional to $d^2/[(d + 2s)\, l_{spr}]$.

Because the energy gap decreases with increasing temperature, this self-heating reduces $V_g$ and makes the $I$-$V$ curves steeper, with larger slope $dI/dV$. In the extreme cases of JJs with high current density (or very poor thermal conductivity) this slope may even become negative, i.e., the voltage decreases with increasing the bias current.

Obviously, the heat removal from the junctions is much better when the chip (or the wafer) is submerged in LHe than on the wafer chuck in vacuum. Even with a perfect thermal contact of the wafer backside with the chuck surface (which can never happen due to wafer bow and surface imperfections), the thermal pass from the junction to the chuck through the 750 μm thick wafer with multiple layers of SiO$_2$ with total thickness of about 3 μm is much longer than through the 0.4 μm thin oxide to the front surface to LHe. Hence, heating effects on the wafer prober should be more pronounced.

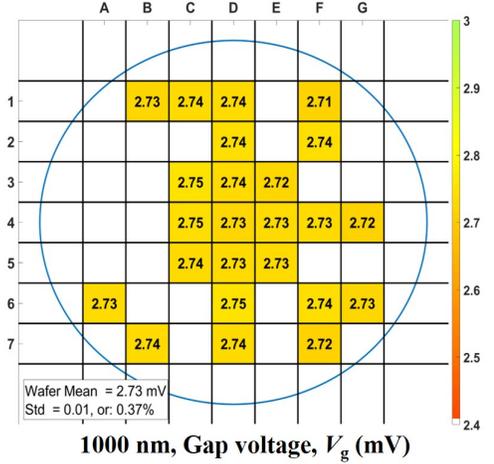

(a)

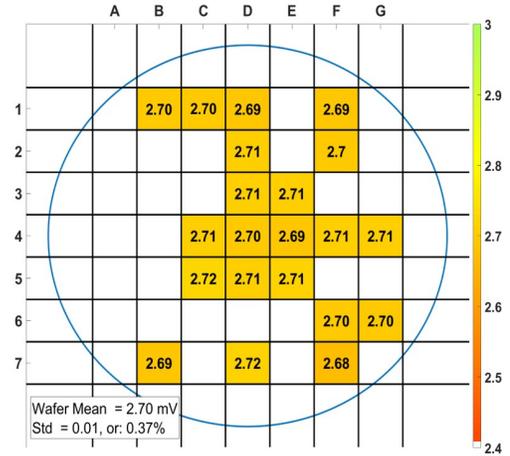

(b)

Fig. 11. Wafermaps of the gap voltage $V_g = 2\Delta/e$ in Nb/Al-AlO$_x$/Nb Josephson junctions with diameters 1000 nm (a) and 1400 nm (b) measured on the same wafer as in Fig. 9. There is an apparent reduction in the gap voltage in the junctions with large sizes with respect to the junctions with small sizes. This is a result of self-heating inside the junctions; see text.

Junction internal temperature $T_J$ after switching to the gap voltage can be estimated using the temperature dependence of the energy gap $\Delta(T)$ in the BSC theory [21]. Energy gap in Nb is known [23] to follow the strong-coupling limit of the BCS theory [24], [24] described by an expression

$$\frac{V_g(T)}{V_g(0)} = \tanh[\frac{T_c}{T}\frac{V_g(T)}{V_g(0)}], \quad (3)$$

The measured value of Nb films critical temperature is $T_c$ =9.1 K, the independently determined zero temperature gap voltage in our junctions is $V_g(0)$ =2.87 mV; see also data in [23]. Then, using (3), the gap voltages of 2.73 mV and 2.7 mV in Fig. 11 correspond to the $T_J$ of, respectively, 4.7 K and 4.9 K at the dissipated power of, respectively, 0.214 µW and 0.420 µW inside 1000-nm and 1400-nm diameter junctions.

Thermal resistance between the junction and the wafer is $R_{th} = \Delta T/P$, where $\Delta T = T_J - T_{wafer}$ and $T_{wafer}$ is the temperature of the wafer clamped to the chuck of the wafer prober. Then, using the above temperatures and heat powers, the wafer temperature during the measurements can be estimated as $T_{wafer}$ = 4.4 K. At this temperature, if junction self-heating were negligible, i.e., $T_J = T_{wafer}$, the expected gap voltage from (3) should be 2.76 mV. This value was indeed observed in the smallest junctions used (500 nm in diameter) at dissipated power $P \approx$ 50 nW, providing additional credence to our estimate of the wafer temperature on the chuck of the wafer prober.

### B. Inductors

Radial distribution of inductance of Nb layer M6 in Fig. 5e-h was attributed in [18] to a slightly larger magnetic field penetration depth, $\lambda$ in Nb film of the signal traces in the central part of the wafer than near the wafer edge. A 3% larger inductance would require $\lambda_1 \approx$ 97 nm in (1).

The inductance enhancement appears to take place in about the same region where $R_N$ of Josephson junctions is increased, but is less pronounced. In microscopic theory of superconductivity, the penetration depth in superconductors with short mean free path of electrons is given by [26], [27]

$$\lambda(T) = \left(\frac{\hbar\rho}{\mu_0 \pi \Delta \tanh\frac{\Delta}{2k_B T}}\right)^{1/2}, \quad (4)$$

where $\rho$ the normal state resistivity, $\hbar$ is the reduced Planck's constant. Hence, $\lambda$ at 4.2 K and inductance of various structures should be larger in those parts of the wafer where resistivity of Nb films is higher than in other parts and/or the energy gap is smaller.

According to the wafermaps in [6], the sheet resistance of Nb films at room temperature is higher in the central part of the wafer, indicating that the resistivity $\rho$ is also higher because the thickness of the films does not substantially decrease radially. At the same time, the wafermaps in Fig. 11 show that the energy gap in (4) does not depend on the location. Hence, the presented inductance wafermaps correlate with the sheet resistance wafermaps in [6] and both point to an increased values of the penetration depth in the central part of the wafer. In this respect, it would be interesting to obtain resistivity wafermaps of Nb films at low temperatures, e.g., at 10 K. Taking the measured value of Nb film resistivity at 10 K of about 5 µΩ cm and the mean energy gap from Fig. 11 of $2\Delta/e$ =2.73 mV, we get from (1) a value $\lambda$(4.2 K) =80 nm which reasonably agrees with the value of 90 nm extracted from the inductance measurements in [17]-[19].

### C. Vias

Distribution of the critical current of vias in Fig. 7 apparently correlates with the distribution of the etch rate in the high-density plasma etching system used for $SiO_2$ etching. Lower etch rate of contact holes in $SiO_2$ near the wafer edge may lead to higher contact resistance between the bottom Nb layer and the next Nb layer deposited over the dielectric into the contact holes to form vias.

## V. CONCLUSION

We performed the first 200-mm wafer-scale testing of superconductor electronics wafers on a cryogenic wafer prober. The testing included all fabrication process control monitors and small integrated circuits for inductance extraction. Using the gap voltage measurements on junctions with different sizes and different power dissipation, we estimated the actual wafer temperature during the measurements as 4.4 K. This value is very close to liquid helium temperature and, thus, all the obtained results can be directly compared with chip-level testing done in LHe. The wafer prober testing revealed distributions of the process parameters (wafermaps) which previously were either unknown or known only qualitatively, and gives indispensable information for process engineers and circuit designers into the process.

The next and more challenging step should be demonstration of wafer-scale testing of complex integrated circuits. We see the main challenges in removing heat from the circuits with substantial power dissipation, e.g., in bias resistors of the standard RSFQ circuits, as well as in providing sufficient suppression of electromagnetic noise.

Results of this work clearly demonstrate advantages of the wafer-scale testing compared to the conventional chip-scale testing involving wafer dicing, individual chip packaging, and cryogenic testing in liquid helium. There is no doubt that full implementation of cryogenic wafer-scale probing in the superconductor fabs will dramatically help advancing superconductor electronics for various applications from traditional sensors and microwave devices to advanced classical computing, to quantum optimization and quantum computing.

## ACKNOWLEDGMENT

We are thankful to Jeff Wang and Xi Lin for assisting with the probe card modification that enabled the testing in this paper. We are grateful to Vlad Bolkhovsky, Ravi Rastogi, and Scott Zarr for overseeing the SFQ5ee process runs and to the entire MIT LL fab team for wafer processing.

This material is based upon work supported by the Under Secretary of Defense for Research and Engineering under Air Force Contract No. FA8702-15-D-0001. Any opinions, findings, conclusions or recommendations expressed in this material are those of the author(s) and do not necessarily reflect the views of the Under Secretary of Defense for Research and Engineering. Delivered to the U.S. Government with Unlimited Rights, as defined in DFARS Part 252.227-7013 or 7014 (Feb 2014). Notwithstanding any copyright notice, U.S. Government

## REFERENCES

[1] S.K. Tolpygo, V. Bolkhovsky, R. Rastogi *et al*., "A 150-nm process node of an eight-Nb-layer fully planarized process for superconductor electronics," Invited presentation Wk1EOr3B-01 at Applied Superconductivity Conference, ASC 2020 Virtual Conference, 25 Oct. − 7 Nov. 2020. Presented 29 Oct. 2020. [Online] Available: https://whova.com/portal/webapp/appli_202010/Agenda/1277628 https://snf.ieeecsc.org/sites/ieeecsc.org/files/documents/snf/abstracts/STP669%20Tolpygo%20invited%20pres.pdf

[2] V.K. Semenov, Y.A. Polyakov, S.K. Tolpygo, "AC-biased shift registers as fabrication process benchmark circuits and flux trapping diagnostic tool," *IEEE Trans. Appl. Supercond.,* vol. 27, no. 4, June 2017, Art. no. 1301409.

[3] M.W. Johnson, P. Bunyk, F. Maibaum, *et al*., "A scalable control system for a superconducting adiabatic quantum optimization processor," *Supercond. Sci. Technol*., vol. 23, Apr. 2010, Art. no. 065004.

[4] A.J. Berkley, M.W. Johnson, P. Bunyk *et al*., "A scalable readout system for a superconducting adiabatic quantum optimization system," *Supercond. Sci. Technol*., vol. 23, Sep. 2010, Art. no. 105014.

[5] Q.P. Herr, J. Osborne, M.J. Soutimore, H. Hearne, R. Selig, J. Vogel, E. Min, V.V. Talanov, and A.Y. Herr, "Reproducible operating margins on a 72800-device digital superconducting chip," *Supercond. Sci. Technol*., vol. 28, Oct .2015, Art. no. 124003.

[6] S.K. Tolpygo, V. Bolkhovsky, R. Rastogi *et al*., "Fabrication processes for superconductor electronics: Current status and new developments," *IEEE Trans. Appl. Supercond*., vol. 29, no. 5, Aug. 2019, Art. no. 1102513.

[7] J.M. Geary and G.P. Vella-Coleiro, "Cryogenic wafer prober for Josephson devices," *IEEE Trans. Magn*., vol. MAG-19, no. 3, pp. 1190-1192, May 1983.

[8] J.M. Kreikebaum, K.P. O'Brien, A. Morvan, and I. Siddiqi, "Improving wafer-scale Josephson junction resistance variation in superconducting quantum coherent circuits," *Supercond. Sci. Technol*., vol. 33, no. 6, 2020, Art. no. 06LT02.

[9] I.W. Haygood, E.R.J. Edwards, A.E. Fox, *et al*., "Characterization of uniformity of Nb/Nb$_x$Si$_{1-x}$/Nb Josephson junctions," *IEEE Trans. Appl. Supercond*. vol. 29, no. 8, Jun. 2019, Art. no. 1103505.

[10] IQ3000 4 K Cryogenic Wafer Prober HPD/FormFactor. 4601 Nautilus Court South, Suite 101. Boulder, CO 80301.

[11] S. K. Tolpygo, V. Bolkhovsky, T. J. Weir, L. M. Johnson, M. A. Gouker, and W. D. Oliver, "Fabrication process and properties of fully planarized deep-submicron Nb/Al-AlO$_x$-Nb Josephson junctions for VLSI circuits," *IEEE Trans. Appl. Supercond*., vol. 25, no. 3, Jun. 2015, Art no. 1101312.

[12] S. K. Tolpygo, V. Bolkhovsky, T. J. Weir *et al*., "Advanced fabrication processes for superconducting very large scale integrated circuits", *IEEE Trans. Appl. Supercond.*, vol. 26, no. 3 , April 2016, Art. no. 1100110

[13] MX500C vacuum transport system. Brooks Automation, Inc. 15 Elizabeth Drive. Chelmsford, MA 01824 U.S.A.

[14] Mag F Probe, Bartington Instruments, Ltd. 5, 8, 10, 11 & 12 Thorney Leys Business Park. Witney, Oxford OX28 4GE. England.

[15] Y. A. Polyakov, V. K. Semenov and S. K. Tolpygo, "3D Active demagnetization of cold magnetic shields," in *IEEE Trans. Appl. Supercond.*, vol. 21, no. 3, pp. 724-727, Jun. 2011.

[16] S. K. Tolpygo, "Superconductor digital electronics: scalability and energy efficiency issues," *Low Temp. Phys. / Fizika Nizkikh Temperatur*, vol. 42, no. 5, pp. 463-485, May 2016.

[17] S.K. Tolpygo, V. Bolkhovsky, T.J. Weir *et al*., "Inductance of circuit structures for the MIT LL superconductor electronics fabrication process with 8 niobium layers," *IEEE Trans. Appl. Supercond*., vol. 25, no. 3, Jun. 2015, Art. no. 1100905.

[18] S.K. Tolpygo, E.B. Golden, T.J. Weir, and V. Bolkhovsky, "Inductance of superconductor integrated circuit features with sizes down to 120 nm," *Supercond. Sci. Technol*., vol. 34, Jun. 2021, Art. no. 085005.

[19] S.K. Tolpygo, E.B. Golden, T.J. Weir, and V. Bolkhovsky, "Mutual and self-inductance in planarized multilayered superconductor integrated circuits: Microstrips, striplines, bends, meanders, ground plane perforations," Arxiv 2110.07799. [On-line] Available: http://arxiv.org/abs/2110.07799

[20] S.K. Tolpygo, V. Bolkhovsky, R. Rastogi *et al*., "Planarized fabrication process with two layers of SIS Josephson junctions and integration of SIS and SFS π-junctions," *IEEE Trans. Appl. Supercond*., vol. 29, no. 5, Aug. 2019, Art. no. 1101208.

[21] J. Bardeen, L.N. Cooper, and J.R. Schrieffer, "Theory of superconductivity," *Phys. Rev*., vol. 108, no. 5, pp. 1175-1204, Dec. 1957.

[22] S.K. Tolpygo, V. Bolkhovsky, S. Zarr, *et al*., "Properties of unshunted and resistively shunted Nb/AlO$_x$-Al/Nb Josephson junctions with critical current densities from 0.1 to 1 mA/μm$^2$," *IEEE Trans. Appl. Supercond*., vol. 27, no. 4, Jun. 2017, Art. no. 1100815.

[23] R.F. Broom, "Some temperature-dependent properties of niobium tunnel junctions," *J. Appl. Phys*., vol. 47, no. 12, pp. 5432-5439, Dec. 1976.

[24] D.J. Thouless, "Strong-coupling limit in the theory of superconductivity," Phys. Rev., vol. 11 no. 5, pp. 1256-1260, Mar. 1960.

[25] J.C. Swihart, "Solution of the BCS integral equation and deviation from the law of corresponding states," *IBM J. Res. Develop*., vol. 6, pp. 14-23, Jan. 1962.

[26] A.A. Abrikosov and L.P. Gor'kov, "On the theory of superconducting alloys. I. The electrodynamics of alloys at absolute zero," *Zh. Eksper. Teor. Fiz*., vol. 35, pp. 1558-1571, Dec. 1958. English translation: *Sov. Phys. JETP*, vol. 8, no. 6, pp. 1090-1098, Jun. 1959.

[27] A.A. Abrikosov and L.P. Gor'kov, "Superconducting alloys at finite temperatures," *Zh. Eksper. Teor. Fiz*., vol. 35, pp. 319-320, Jan. 1959. English translation: *Sov. Phys. JETP*, vol. 9, no. 1, pp. 220-221, 1959.